\documentclass[letter]{aa} 

\usepackage{placeins}
\usepackage{natbib,twoopt}
\usepackage{upgreek}   
\usepackage[breaklinks=true]{hyperref} 
\usepackage{color}
\bibpunct{(}{)}{;}{a}{}{,}             

\usepackage{graphicx}
\usepackage{txfonts}
\graphicspath{{./}{figures/}}

\begin{document}

    \title{A trio of month long flares in the nova-like variable V704 And}

    \author{Gabriella Zsidi\inst{1,2,3}
          \and
          C. J. Nixon\inst{1}
          \and
          T. Naylor\inst{4}
          \and
          J. E. Pringle\inst{5}
          \and
          K. L. Page\inst{6}
          }

    \institute{School of Physics and Astronomy, Sir William Henry Bragg Building, Woodhouse Ln.,
    University of Leeds, Leeds LS2 9JT, UK\\               \email{g.zsidi@leeds.ac.uk}
    \and
    Konkoly Observatory, Research Centre for Astronomy and Earth Sciences, Eötvös Loránd Research Network (ELKH), Konkoly-Thege Miklós út 15-17, 1121 Budapest, Hungary
    \and 
    CSFK, MTA Centre of Excellence, Budapest, Konkoly Thege Mikl\'os \'ut 15-17, 1121, Hungary
    \and
    School of Physics and Astronomy, University of Exeter, Stocker Road, Exeter EX4 4QL, UK  
    \and
    Institute of Astronomy, University of Cambridge, Madingley Road, Cambridge, CB3 0HA, UK
    \and
    School of Physics and Astronomy, University of Leicester, Leicester LE1 7RH, UK
    }

    \date{Received \today}

    \abstract{We present the discovery of an unusual set of flares in the nova-like variable V704 And. Using data from AAVSO, ASAS-SN, and ZTF, of the nova-like variable V704 And, we have discovered a trio of brightening events that occurred during the high state. These events elevate the optical brightness of the source from $\sim13.5$ magnitude to $\sim12.5$ magnitude. The events last for roughly a month, and exhibit the unusual shape of a slow rise and faster decay. Just after the third event we obtained data from regular monitoring with Swift, although by this time the flares had ceased and the source returned to its pre-flare level of activity in the high-state. The Swift observations confirm that during the high-state the source is detectable in the X-rays, and provide simultaneous UV and optical fluxes. As the source is already in the high-state prior to the flares, and thus the disc is expected to already be in the high-viscosity state, we conclude that the driver of the variations must be changes in the mass transfer rate from the companion star and we discuss possible mechanisms for such short-timescale mass transfer variations to occur.
    }


   \maketitle
%

\section{Introduction}
Cataclysmic variables (CVs) are close interacting binary systems, in which a white dwarf is accreting material from a companion via Roche-lobe overflow. 
In the case of weakly or non-magnetic CVs, the material transferred from the secondary component forms an accretion disk around primary.
These systems may be classified based on their photometric and spectroscopic characteristics \citep{warner1995}. For example, dwarf novae (DNe) show recurring accretion related outbursts, and sometimes superoutbutsts, lasting from a few days to $\sim20$ days. The nova-like variables (NLs), defined as those that do not display nova or dwarf-nova outbursts, remain in the high accretion, therefore bright photometric state, for a prolonged time.

The VY~Sculptoris stars (also called ``antidwarf novae'') are an interesting subclass of NLs, whose bright photometric states are occasionally interrupted by rapid drops of brightness (low states).
The low states are believed to be caused by decreased mass transfer from the companion, which can eventually lead to the disappearance of the disc \citep{hameury2002}.
However, it is argued that sometimes the accretion disc does not disappear completely during these periods \citep{schmidtobreick2018}.
The exact process causing the reduced mass transfer is still unknown, but one mechanism that was suggested by previous works is starspots on the secondary blocking the L1 point and hence suppressing the mass transfer \citep{livio1994}.

V704~Andromedae, hereafter V704~And, is a poorly studied NL; its variable nature was discovered only roughly two decades ago by \cite{dahlmark1999}, who reported brightness changes in the $V$ band between 14.8 and 12.6\,mag.
The source was further studied in the early 2000s by \cite{papadaki2006} within a photometric survey of not well known CVs, mainly focusing on nova-like variables.
Based on optical photometric measurements between 2002 and 2005, they found that V704~And showed fading episodes in 2003 and 2005.
Long-term historical light curves confirm the occasional fading of the system, which suggests that V704~And belongs to the VY~Scl subclass of the nova-like variables, and the optical study by \cite{weil2018} supports this argument. \cite{bruch2022} further investigated the system and measured an orbital period of $P_{\textrm{orb}} = 0.15424(3)$\,days (3.7\,h).
This is within the typical range of orbital periods for VY~Scl nova-like variables.
Furthermore, they estimated the spectral type of the secondary to be M3--M4 using optical spectra obtained during a low state.
\cite{bruch2022} revealed a strong negative superhump in the V704~And system by analysing the TESS satellite light curves with a period of $P_{\textrm{nSH}} = 0.14772(3)$\,days.
The system was also included in the Gaia survey, which placed it at a distance of 410.4$^{+6.3}_{-5.3}$\,pc \citep{bailerjones2021}.

\section{Observations}

   \begin{figure*}[t!]
   \centering
   \includegraphics[width=\textwidth]{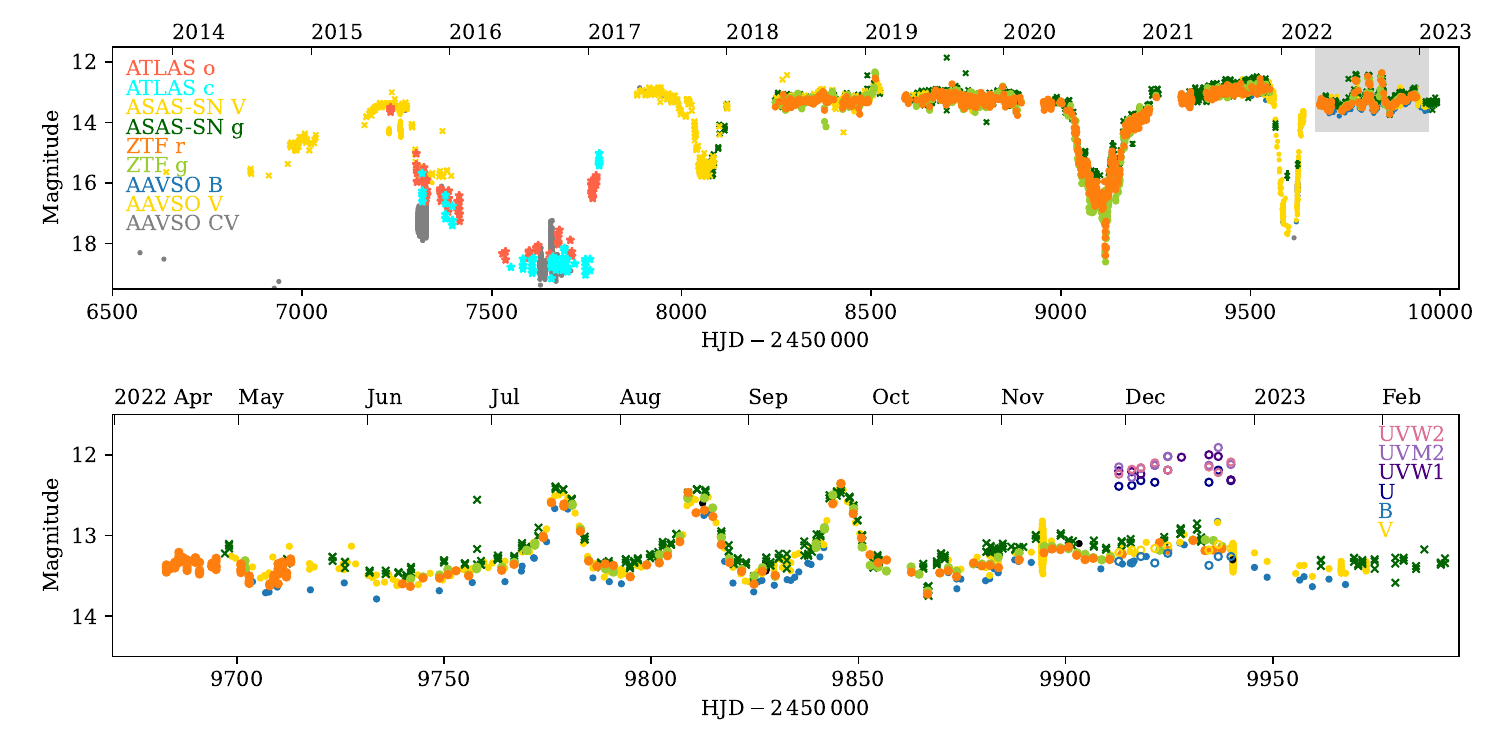}
   \caption{Light curves of V704~And. \textit{Top panel:} Optical light curves of V704~And. The yellow and dark green x-symbols show the ASAS-SN $V$ and $g$ band measurements, the orange and the light green filled circles indicate the ZTF $r$ and $g$ band observations; and yellow, blue and grey circles mark the AAVSO B, V band and unfiltered CCD (CV) measurements, respectively. The cyan and dark orange star symbols are measurements from the ATLAS survey with the $c$ and $o$ filters, respectively. We indicated, with the grey shaded area, the period which is shown in the bottom panel. 
   We note that the ATLAS measurements from 2016 suggest that the system was as faint as $\sim18$\,mag during this period, however, the ASAS-SN data indicated a brightness around 16\,mag. Since the sensitivity of ATLAS observations is higher, and the ASAS-SN measurements from this period include several upper limit measurements, we excluded from the figure those ASAS-SN data which were fainter than 16\,mag.
   \textit{Bottom panel:} Optical and UV measurements obtained since 2022 April (grey shaded area in the top panel). The empty circles show the Swift optical and UV data. Yellow, blue and black filled circles are the AAVSO $V$, $B$ and visual band observations, respectively.}
    \label{fig:v704and_lc}%
    \end{figure*}

In order to analyse the long-term behaviour of V704~And, we inspected photometric data from public surveys.
We studied the data obtained by ZTF \citep[][]{bellm2019}, which aims to observe the northern sky with a $\sim$2-day cadence. 
We downloaded the measurements for V704~And from their archive\footnote{\url{https://irsa.ipac.caltech.edu/Missions/ztf.html}}, which provided photometric measurements between 2018 March and 2023 January in the Palomar/ZTF $g$ and $r$ bands for the latest data release (DR16).
In order to remove bad-quality observations, we applied the ``catflags=0'' filter of the catalogue.
Johnson $V$ and Sloan $g$ band photometric observations are also available from the ASAS-SN survey. 
The ASAS-SN team performs aperture photometry using a 2-pixel aperture radius, and the reduced data are available from their online catalogue\footnote{\url{https://asas-sn.osu.edu/}}.
$V$ band measurements were obtained between 2013 December and 2018 August, whereas the $g$ band observations are available since 2017 December \citep{kochanek2017, shappee2014}.
Despite both surveys providing $g$ band data, we analysed them separately as the sensitivity of the two surveys differ: the ASAS-SN is sensitive up to $\sim$17\,mag, whereas the ZTF is accurate up to about 20.8\,mag. 
We also complemented our dataset with observations from the ATLAS survey \citep{heinze2018}, which provided measurements in ``cyan'' and ``orange'' filters\footnote{The ``cyan'' ($c$) filter covers the 420–650 nm wavelength range, and the ``orange'' ($o$) filter covers the 560–820 nm band.} between 2015 and 2017, which have a sensitivity up to $\sim19$\,mag.
Furthermore, we examined the AAVSO optical light curves of V704~And, which date back to 1966 and allow to study the historic behaviour of the system.
However, we must note that the early AAVSO observations for V704~And are rather incomplete, and they either did not provide information on the filter used, or applied a ``visual'' band. 
The more recent measurements often provide Johnson $B$ or $V$ band observations besides the visual ones\footnote{\url{https://www.aavso.org/filters}}.


We also have been monitoring V704~And with the Ultraviolet/Optical Telescope (UVOT) instrument of the Swift satellite between 2022 November 29 and December 26 with a three-day cadence (ToO ID 18163, PI: G. Zsidi). 
We used the ``0x30ed'' mode of UVOT, which provided photometric measurements in the $V$, $B$, and $U$ optical and the $UVW1$, $UVM2$, and $UVW2$ UV filters\footnote{Central wavelengths of the UVOT filters ($\AA$): $V$ -- 5468, $B$ -- 4392, $U$ -- 3465, $UVW1$ --  2600, $UVM2$ -- 2246, $UVW2$ --  1928.}.
However, due to scheduling conflict, one observation was not executed, and the one on 2022 December 14 was carried out only in the $UVW1$ filter.
We reduced the UVOT data using the version 6.31.1 of the \textsc{heasoft} software. 
All UV and optical light curves are displayed in Fig.~\ref{fig:v704and_lc}.

\section{Results}\label{sect:results}
The brightness of V704~And varies around 13\,mag in the high state but this is occasionally interrupted by low state periods, during which the brightness of the system decreases by $3-5$\,mag.
According to the most recent optical ASAS-SN and the ZTF light curves (Fig.~\ref{fig:v704and_lc}, top panel), V704~And underwent three low state periods during the last six years (i.e. since 2017), each of which lasted about six months. 
Before 2017, the system exhibited an inverse behaviour to some extent, as it spent a longer period of time in the low brightness state.
This behaviour was characteristic of the system since the mid-2000s, as revealed by the AAVSO light curves (Fig.~\ref{fig:v704and_aavso_lc}).
Despite the fact that the AAVSO observations date back to 1966, the early measurements were rather incomplete, therefore, we cannot make far reaching conclusions about the historical behaviour of the system.
The available data suggest that V704 And spent significant time in the high state between 1966 and 1995, but intervening drops into the low state during this time cannot be ruled out.

\subsection{The trio of flares in the latest high state}
The most recent low state of V704~And finished around March/April in 2022. The source initially returned to a regular high state for a few months, albeit at a slightly lower flux level than before the latest minimum ($g\approx13.5$\,mag rather than $g\approx12.8$\,mag prior). However, the system then began to show peculiar outburst-like events, which we call flares to distinguish from ordinary CV outbursts, with amplitudes of $\sim$1 mag and lasting for about a month each.
The bottom panel of Fig.~\ref{fig:v704and_lc} shows an enlarged excerpt of the light curve highlighting these events.
The outbursts seem to have an asymmetric evolution: it takes $18-20$\,days to reach the maximum, and $10-12$\,days to return to the initial brightness level\footnote{We also note that around $\textrm{HJD}-2450000=8500$, V704~And exhibited a small amplitude brightening event. However, due to the sparseness of the data during that period, it is not possible to tell with certainty whether this event was similar to the trio of flares in 2022.}.

The light curves of the flares obtained with different filters are remarkably similar, we did not detect any noticeable time delay in reaching the maximum brightness at different wavelengths.
This might be caused by two factors.
One is that, due to the cadence of the light curves, the peaks of the flares are not well defined in all filters.
The other cause could be that the available optical filters indeed probe the same region of the system; therefore, no time delay is expected in these observations.

It is instructive to compare the outbursts of V704~And with the dwarf novae outbursts.
The two types of outbursts might look similar in amplitude and in length, but the dwarf nova outbursts normally exhibit an asymmetry with an opposite trend, i.e. they show faster rise than decay \citep{cannizzo1986}.
Furthermore, the increase of the far-UV flux can be delayed compared to the optical.
This second aspect, unfortunately, cannot be probed for V704~And, as we have only optical light curves during the flares.


\subsection{Colour changes}
\begin{figure}[t!]
   \centering
   \includegraphics[width=0.65\linewidth]{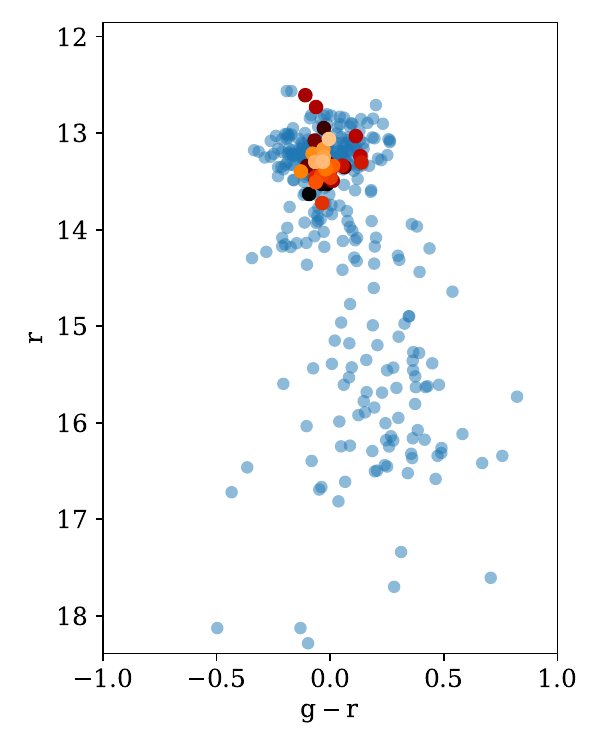}
   \caption{Optical colour-magnitude diagram based on the ZTF measurements. The blue filled circles show the entire dataset, whereas the orange ones cover only the period between $\textrm{JD}-2450000 = 9740\ \textrm{and}\ 9890$, i.e. when the three flares occurred. The measured errors are smaller than or equal to the symbol size.}
   \label{fig:v704and_cmd}%
   \end{figure}

The multi-filter photometric observations allow us to examine the colour variations of V704~And.
The ZTF survey has long enough coverage to study the changes in the system on yearly timescale.
We constructed the colour-magnitude diagram using the simultaneous $g$ and $r$ band measurements (Fig.~\ref{fig:v704and_cmd}).
Our results show that when the object is in the low and fainter state, the data points cover a wider range in the colour space.
However, as the system brightens, the overall colour is more concentrated on the blue side.

It is interesting to separate those observations when the system underwent the three flares in order to investigate their nature.
For this reason, we highlighted the data points that cover the period between $\textrm{HJD} - 2\,450\,000 = 9740$ and $9890$ with orange symbols.
Furthermore, we indicated the earlier measurements with darker shades and the later ones with lighter in order to investigate the temporal evolution.
We found that no significant colour change was observed when V704~And brightened by $\sim$1\,mag.
However, we must note that the simultaneous $g$ and $r$ band data do not cover the flares entirely.

The Swift measurements provide information on the colour of the high state after the trio of flares.
During this time, the brightness of the system did not change substantially.
Fig.~\ref{fig:v704and_swift_cmd} shows, that no significant colour change was detected in the optical filters.
The most apparent evolution is observed at the shortest wavelengths ($UVM2$ and $UVW2$ filters), in which the system becomes redder as it brightens.

\section{Discussion}\label{sect:discussion}

V704 And is a nova-like variable star that typically displays a steady high-state brightness corresponding to $g\approx13-13.5$\,mag, with aperiodic low states lasting several months. In the high-state the accretion disc shows no evidence for DNe type outbursts, and thus the whole disc is expected to be fully ionised with well-established hydro-magnetic viscosity. In similar systems, primarily the DNe in outburst, the disc viscosity parameter $\alpha$ \citep{shakura1973} is found to be high with $\alpha \approx 0.2-0.4$ \citep{martin2019}. As the disc is already in the high-state for the months preceding the trio of flares it is unlikely that these flares are caused by variations in $\alpha$. Instead, they are likely to be caused by variations in the accretion rate through the disc, which is controlled by the mass supply rate from the companion star. To determine whether the duration of the flares is consistent with varying mass supply to the outer disc regions, we compute the viscous timescale in V704 And and then we discuss possible mechanisms to induce a change to the mass transfer rate in this system.

\subsection{Viscous timescale in V704 And}
The timescale on which the central accretion rate responds to changes in the mass transfer rate from the companion star is the viscous timescale given by $t_\nu = R^2/\nu$. Here $\nu$ is the \cite{shakura1973} kinematic viscosity and $R$ is the radius in the disc at which the mass transferred from the companion star is added. For our estimate, we take this to be the outer disc edge where viscous spreading of the disc is impeded by tides from the companion star. To evaluate $R$ and $\nu$ we require estimates for the system parameters, e.g. $M_1$, $M_2$, $P_{\rm orb}$, and $\dot{M}$.

The estimates of the system parameters can be uncertain, therefore, we only summarise here the main steps of the calculation.
Our detailed calculation and the estimates of the system parameters can be found in Appendix~\ref{app:timescale}.
By using $P_{\rm orb} = 0.15424(3)$\,days, $q = 0.34$, $M_2 = 0.3\,M_{\odot}$, and $M_1 = 0.9\,M_{\odot}$, we have that the binary separation is $a = 8.9\times 10^{10}$\,cm, and therefore the outer disc radius is $R \approx 2.3\times 10^{10}$\,cm. 
We also assume a typical white dwarf radius of $R=0.01$\,$R_{\odot}$, and take $\alpha = 0.3$. Then using equation 5.50 in \cite{frank2002}, we estimate the viscous timescale through the disc to be 9.6\,days for an accretion rate of $5\times 10^{-9}M_\odot$/yr, and 7.7\,days for an accretion rate of $10^{-8}M_\odot$/yr. 
We therefore conclude that the viscous timescale is approximately a week, and maybe up to 10 days. This timescale is consistent with the breadth of the flares, and also consistent with the viscous timescale inferred from measurements of the decay of DNe light curves \citep[e.g.][]{cannizzo1994}.

\subsection{Origin of the flares}\label{sect:origin}

The origin of the flares is puzzling. 
At the optical wavelength range, the accretion disc is the most dominant component of a NL system in high state \citep{pringle1985}, however, disc instability typical of DNe is an unlikely cause of the observed flares of V704~And, as their characteristics differ from DN outbursts.
We propose that variable mass transfer from the secondary is responsible for inducing the flares.
Nonetheless, VY~Scl type variables are expected to have the highest mass transfer rate during the high state \citep{warner1987}.
For this reason, there has to be an additional mechanism that induces the increase of the mass transfer rate.
Here, we discuss a few mechanisms to achieve this.

Variable mass transfer rate can be achieved by the variations of the radius of the secondary \citep{warner1988, king1995}.
\cite{king1995} proposes that variable irradiation by the accreting component may induce changes in the radius of the Roche lobe-filling star, which causes a limit-cycle variation.
However, the timescale of this is significantly longer ($10^5 - 10^6$\,yr) than the variations we see here.
Another mechanism that could contribute to the variation of the radius of the secondary component is solar-type magnetic cycles.
\cite{warner1988} report such quasiperiodic variations in CVs but these also appear on several-years timescale.

Magnetic activity, at the same time, can have short-term effects as well \citep{baliunas1985}.
The light curves of main sequence stars often reveal the presence of star spots and flares, and as the secondaries of CVs are main sequence stars, stellar activity is expected in these systems as well.
Indeed, e.g. \cite{webb2002} have found spectroscopic evidence for starspots on the secondary star of SS~Cyg, and \cite{dunford2012} found a high-latitude starspot on RU~Peg with Roche tomography.
Flares or coronal mass ejections of the secondaries of a CV may induce additional large amounts of matter, however, we cannot directly test this theory for V704~And, as the optical light curves of CVs are dominated by the emission from the disc, which veils the variability of any stellar activity.

\subsection{Comparison with other VY~Scl type systems}
\label{sect:comparison}

We have compiled a list of VY Scl type NLs and examined their short- and long-term light curves based on AAVSO, ZTF and ASAS-SN observations with the aim of looking for any similar behaviour to the one observed in V704~And.
The majority of these objects are taken from Table~4.1 of \cite{warner1995}, which we complemented with a few additional systems which were identified as VY~Scl type NLs since that table was compiled.
See our complete list of the inspected objects in Table~\ref{table:vyscl_list}.
We found that most of these systems behave as typical VY~Scl type objects with high states occasionally interrupted by low states.
We identified only a few systems, which show any kind of outburst-like events (see Fig.~\ref{fig:vyscl_gallery}).
For example, the light curves of HV~And, BZ~Cam, LQ~Peg, and HS~0506+7725 each reveal a small amplitude brightening, but these events show more complex structure than those of V704~And.
In contrast, V1082~Sgr and VZ~Scl show more stochastic brightness variations.

When individual outbursts are examined, some VY~Scl type systems show superficial similarities to the flares of V704~And.
Certain brightening events in the light curves of FY~Per, CM~Del, or LN~UMa seem to have similar duration and/or amplitude to the flares seen in the light curves of V704~And. However, there are significant differences between V704~And and these other NLs. Firstly, the shape, amplitude, and duration of the brightening events of the latter vary from outburst to outburst, whereas the three flares of V704~And are remarkably similar. This is also supported by the high cadence TESS light curve of FY~Per (Fig.~\ref{fig:fyper_tess}) which shows a faster rise than decay, i.e. the opposite of the flares in V704~And.
Secondly, the long-term light curves reveal that the outbursts can be continuously present in the light curves of FY~Per and CM~Del over timescales of decades (see Fig.~\ref{fig:longterm_lc}), which is best revealed by the higher cadence ASAS-SN and ZTF data; in contrast V704~And displayed three identical flares over a period of three months.

Among the sources that show features which are superficially similar to the V704~And flares, LN~UMa is the closest match of any system to V704~And. LN~UMa shows alternating low states and high states typical of a VY~Scl type object \citep{hillwig1998}. However, during the last decade the lightcurve of LN~UMa appears significantly more variable than the regular high states of V704~And. In particular, the high states of LN~UMa are often interrupted by a series of outburst-like events, which can remain present for an extended period of time. This behaviour is not characteristic of VY~Scl type objects, and to our knowledge has not been reported for this source in the literature before. It is tempting to assign to the outbursts seen in the LN~UMa lightcurve the same physical mechanism as for the flares seen in V704~And. However, we note that the accretion rate of LN~UMa is likely to be very close to the critical accretion rate for disc instability; using the 974$^{+22}_{-21}$\,pc distance measured by Gaia \citep{bailerjones2021} and the correlation between the absolute magnitude of the disc and the accretion rate \citep[see eq.~7 in][]{puebla2007} yields an estimate for the accretion rate that (depending on the system inclination) can be above or below the critical accretion rate. As such it is not currently possible to decide with any certainty whether LN~UMa is above, on, or below the critical accretion rate, and thus we cannot be confident which physical effect is responsible for the variability in the LN~UMa lightcurve.

Some of the above-mentioned NLs (e.g. FY~Per, V794~Aql) are known to exhibit quasi-periodic stunted outbursts\footnote{The cadence of the lightcurve for V794~Aql is insufficient to accurately characterise the shape of the outbursts (see Fig.~\ref{fig:vyscl_gallery})}, which could be caused by a variety of mechanisms, which include mass transfer modulations, DN type outburst, DN type outbursts in a truncated disc, or Z~Cam-like outbursts \citep{honeycutt1998}.
\cite{honeycutt2001} argues that among these, the accretion disc instability is the favoured driving mechanism.
More recent works show strong indications that some of the stunted outbursts are caused by mass transfer variations \citep[see e.g.][]{robertson2018}. However, these latter authors remain uncertain of the nature of the stunted outbursts due to discrepancies between their findings, as some of the results could be explained by both mass transfer variations and accretion disc instability related to the Z~Cam phenomenon.
The contradiction regarding the origin of the stunted outburst is also discussed in the review by \cite{hameury2020}.

As some of the outbursts in the repeating systems (particularly FY~Per, LN~UMa and CM Del) show some similarity to the three isolated flares in V704~And it is possible that the same mechanism is the driver of both phenomena, including the stunted outbursts. If these phenomena are all caused by the same mechanism then, for the reasons argued in Sect.~\ref{sect:origin}, this leads to the conclusion that the stunted outbursts and the variability in e.g. CM~Del are driven by mass transfer variations rather than disc instability. However, we emphasise that there are significant differences between the three flares seen in V704~And and the variability we have found in other systems; notably the typical shape and number of the events. 
For this reason, we conclude that the flares reported here in V704 And are more likely to be physically distinct and therefore could in principle be driven by a different mechanism.

All in all, the flare nature observed in V704~And is extraordinary; it not only differs from the dwarf nova outbursts, but no similar behaviour has previously been reported in other VY~Scl type NLs.
The light curves of V704~And show typical VY~Scl type behaviour with a constant brightness in the high state occasionally interrupted by short periods of low states.
As the system recovered from the latest low state in 2022 it re-established its regular high state. However, instead of remaining at the usual high state brightness level, it exhibited three flares and only then it returned to the usual plateau of the high state (see lower panel of Fig.~\ref{fig:v704and_lc}).
This makes V704~And the first clear example of a system which exhibits flares on the plateau of the high state.

\section{Conclusions}\label{sect:conclusions}

We report here the discovery of a trio of flares occurring during the high state of the VY~Scl type variable V704~And, behaviour which has never been reported for a nova-like system before.
The observed phenomenon also differs from the dwarf nova outbursts, therefore, it is unlikely to be caused by disc instability.
Here, we propose that the flares are caused by increased mass transfer from the secondary and suggest a few mechanisms, such as stellar activity of the secondary, that could induce it even when the system is already in the high state. 
Future observations of V704~And during a prolonged minimum would be desirable, and would allow for a better understanding of the nature and activity of the secondary star.

\begin{acknowledgements}
We thank the referee for useful comments which helped to improve the manuscript.
We thank Phil Evans for providing technical support with the installation of the Swift data analysis software. This work made use of data supplied by the UK Swift Science Data Centre at the University of Leicester. GZs and CJN acknowledge support from the Leverhulme Trust (grant number RPG-2021-380). CJN acknowledges support from the Science and Technology Facilities Council (grant number ST/Y000544/1).
KLP acknowledges the support from the UK Space Agency.
We acknowledge with thanks the variable star observations from the AAVSO International Database contributed by observers worldwide and used in this research. 
\end{acknowledgements}

\bibliographystyle{aa}
\bibliography{bibliography_v704and.bib}

\begin{appendix}
\section{AAVSO light curve of V704~And}
\label{app:AAVSO}
\begin{figure}[ht!]
\centering
\includegraphics[width=\linewidth]{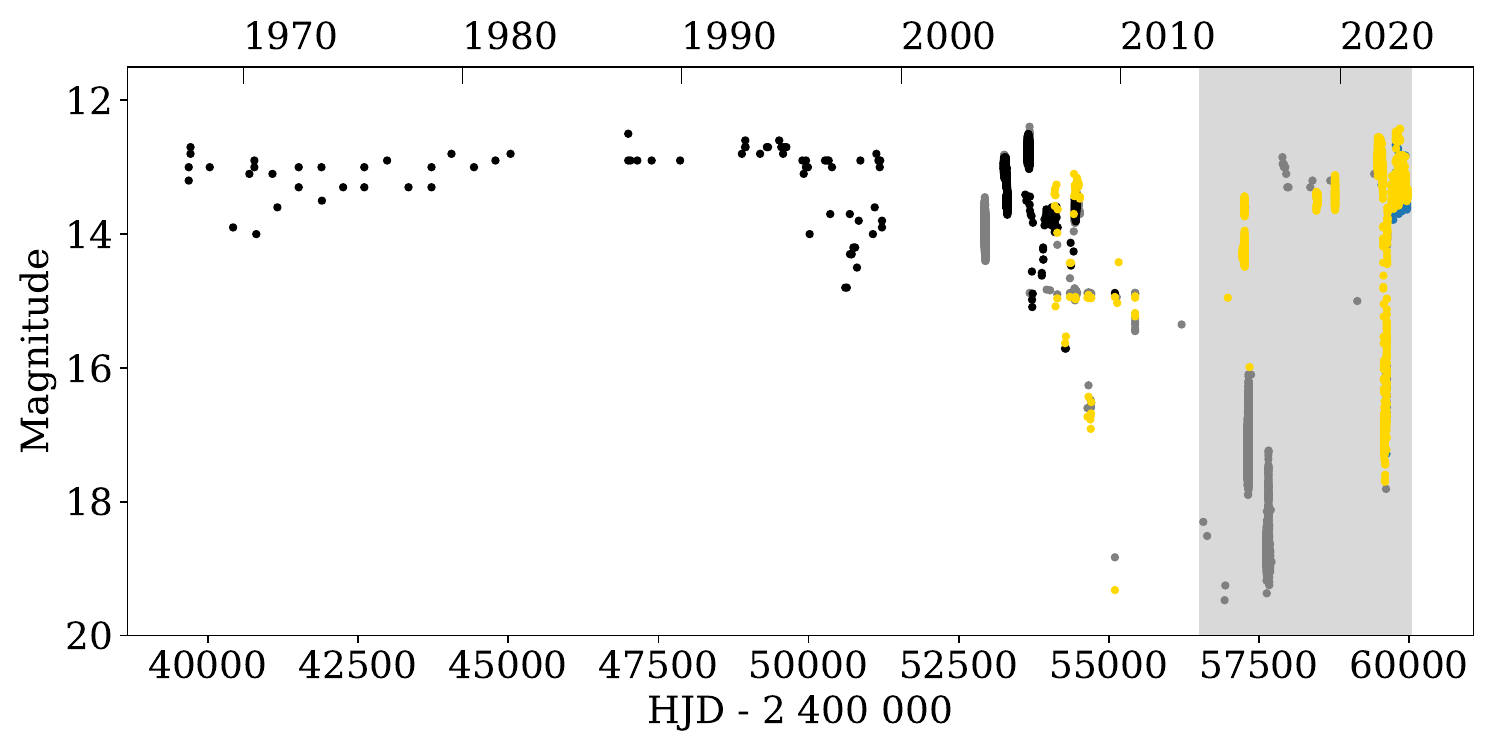}
\caption{AAVSO optical light curves of V704~And. The yellow, blue, black, and grey circles show the B, V, visual, and unfiltered CCD measurements, respectively. We indicated with the grey shaded area the period, which is shown in the top panel of Fig.~\ref{fig:v704and_lc}.}
\label{fig:v704and_aavso_lc}
\end{figure}

AAVSO has been carrying out observations of V704~And since 1966. 
We show the entire available AAVSO light curve for V704~And in Fig.~\ref{fig:v704and_aavso_lc}.

\section{Swift X-ray observations}
\label{app:xray}

   \begin{figure}[ht!]
   \centering
   \includegraphics[width=\linewidth]{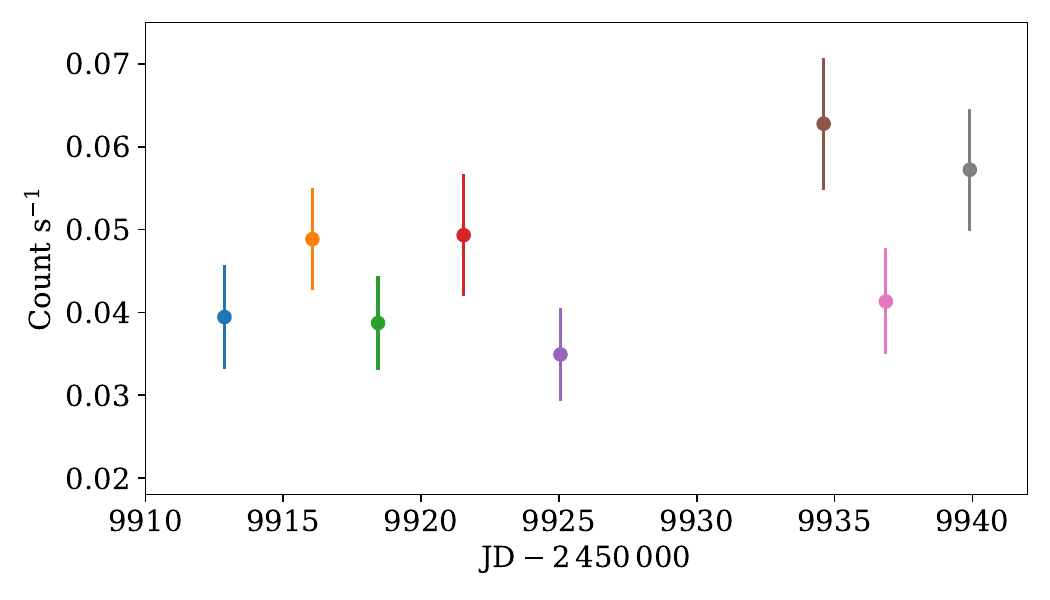}
   \includegraphics[width=\linewidth]{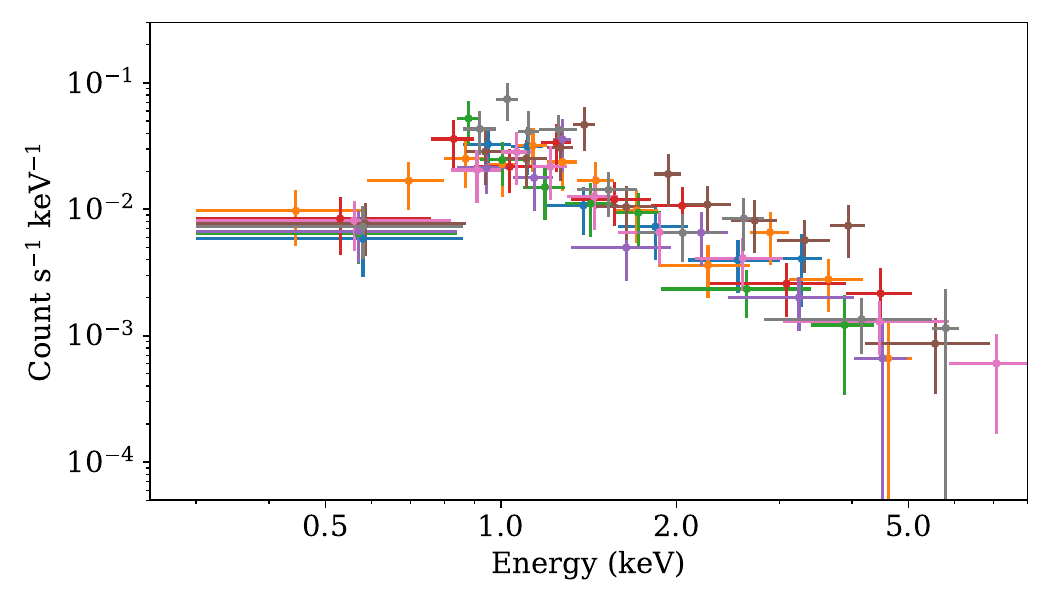}
   \caption{\textit{Top:} XRT light curve. \textit{Bottom:} XRT spectra. The different colours  correspond to the spectra taken at the different epochs, as in the top panel.}
   \label{fig:v704and_swift_xrt}%
   \end{figure}

The Swift satellite also provided X-ray measurements obtained by the X-Ray Telescope (XRT; operates in the $0.3 - 10$\,keV range) contemporaneously with the UVOT observations.
The X-ray data were obtained in photon counting mode for about 1.5\,ks each. 
The spectra were created using the XRT online product generator \citep{evans2009}, which provided the X-ray count rate and spectrum for each epoch.
We show the XRT light curve in Fig.~\ref{fig:v704and_swift_xrt}, which does not reveal high variability during the time of the observations.
The brightness of the source fluctuates around 0.05\,counts\,s$^{-1}$.
We show the X-ray spectra in the bottom panel of Fig.~\ref{fig:v704and_swift_xrt}.
The different colours correspond to the spectra taken at the different epochs.

\section{Confirmation of the viscous timescale in V704 And}
\label{app:timescale}

In order to confirm the viscous timescale of V704~And we require estimates for the system parameters, e.g. $M_1$, $M_2$, $P_{\rm orb}$, and $\dot{M}$. 
The orbital period has been measured by \cite{bruch2022} as $P_{\rm orb} = 0.15424(3)$\,days. To estimate the primary and secondary masses we use the superhump period to determine the mass ratio, and the spectral type of the secondary to determine $M_2$. \cite{wood2009} presents a relationship between the mass ratio $q=M_2/M_1$, the orbital period, and the negative superhump period deficit $\varepsilon_{-}= (P_{\textrm{nSH}} - P_{\textrm{orb}})/P_{\textrm{orb}}$.
Using the $P_{\textrm{orb}}$ and $P_{\textrm{nSH}}$ values determined by \cite{bruch2022}, the mass ratio for the system is $q = 0.34$. \cite{weil2018} estimate the spectral type of the secondary to be M3$-$M4, and based on this we can estimate the mass of the secondary to be around 0.3$M_{\odot}$\footnote{\url{http://www.pas.rochester.edu/~emamajek/EEM_dwarf_UBVIJHK_colors_Teff.txt}} \citep{pecaut2013}.\footnote{This is consistent with the mass estimated from the mass-period relation for CVs which yields $M_2 \approx 0.33\,M_\odot$ \citep[e.g. eq. 2.100 of][]{warner1995}.} Combined with the mass ratio of $q=0.34$, this means that the white dwarf mass is $M_1 = 0.9\,M_{\odot}$.

To estimate the accretion rate through the disc we use the weak correlation between $P_{\textrm{orb}}$ and $\dot{M}$ \citep[][see also \citealt{warner1987}]{puebla2007}, which suggests ${\dot M} \approx 5 \times 10^{-9}$\,$M_{\odot}/\textrm{yr}$. As this is a weak correlation, we expect that values of up to ${\dot M} \approx 10^{-8}\,M_\odot$/yr are plausible, so we provide a range of timescales based on a range of accretion rates below. We note that these accretion rates result in a surface temperature of $\approx (2-3)\times 10^4$\,K in the outer disc regions, which is sufficient to ensure that the source remains in the high-state without exhibiting dwarf nova like outbursts.
Furthermore, the absolute magnitude of V704~And is comparable with other members of the VY~Scl type variables, which also supports the argument that the high state of V704~And is similar to that of other prolonged high states of this class.

\cite{frank2002} provide the outer radius of the disc due to tides as $R \approx 0.9 R_{\rm RL}$ (their equation 5.122), where $R_{\rm RL}$ is the Roche lobe radius given by \citep{eggleton1983}
\begin{equation}
\frac{R_{\rm RL}}{a} = \frac{0.49q^{2/3}}{0.6q^{2/3} + \ln(1 + q^{1/3})}\,
\end{equation}
where $a$ is the binary separation. From the period and masses given above, we have that the binary separation is $a = 8.9\times 10^{10}$\,cm, and therefore the outer disc radius is $R \approx 2.3\times 10^{10}$\,cm. 

Finally, we assume a typical white dwarf radius of $R=0.01$\,$R_{\odot}$, and take $\alpha = 0.3$. Then using equation 5.50 in \cite{frank2002} we estimate the viscous timescale through the disc to be 9.6\,days for an accretion rate of $5\times 10^{-9}M_\odot$/yr, and 7.7\,days for an accretion rate of $10^{-8}M_\odot$/yr. We therefore conclude that the viscous timescale is approximately a week, and maybe up to 10 days. 
The system parameter estimates also allowed us to determine $\dot{M}_{crit} = 1.26 \times 10^{-9}M_\odot/\textrm{yr}$ for V704~And using equation~13 in \citet[][see \citealt{lasota2008}]{hameury2020}, which is lower than $\dot{M}$, i.e. the system is expected to be in the stable regime.

\section{Other VY~Scl type nova-like variables}
\label{app:vyscl}
\begin{figure*}
\centering
\includegraphics[width=\textwidth]{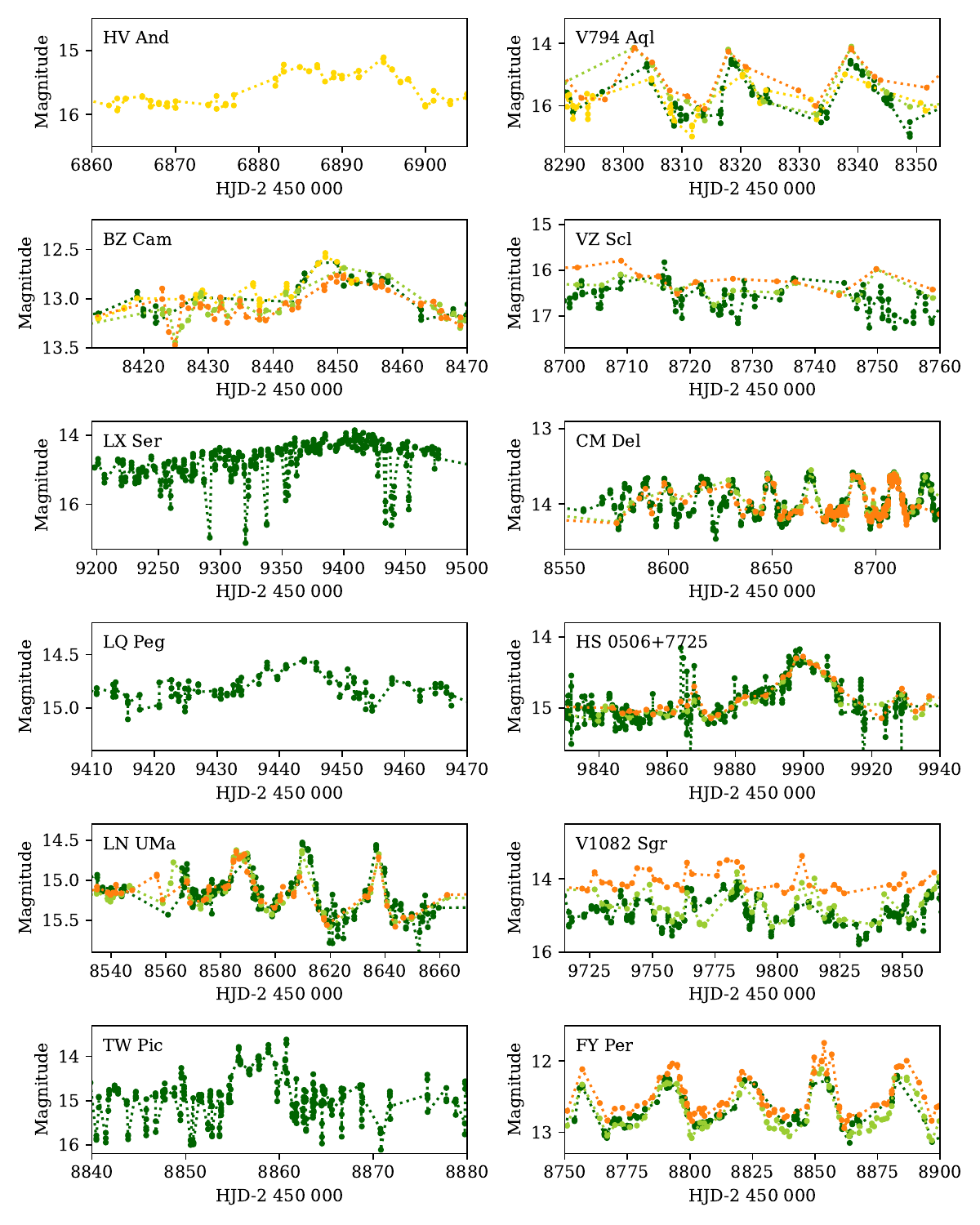}
\caption{Excerpts of the optical light curves of a few VY~Scl type nova-like variables. The yellow and the dark green symbols show the ASAS-SN $V$ and $g$ band observations, respectively, and the orange and the light green filled circles indicate the ZTF $r$ and $g$ band observations. The names of the systems are displayed in the top left corner of each panel.}
\label{fig:vyscl_gallery}
\end{figure*}

We have examined a list of VY~Scl type nova-like variables in order to find similar outbursting behaviour to the one observed in V704~And.
In general, none of these objects show outbursts with similar nature, i.e. slower rise and faster decay.
Here, we give comments on a few individual objects, which exhibit some kind of outburst-like event, and show the corresponding excerpts of their light curves in Fig.~\ref{fig:vyscl_gallery}
\paragraph{HV And} 
The source exhibits one brightening event which lasts for about 30\,days but has small amplitude and asymmetric shape.
\paragraph{V794 Aql} The system shows $\sim$2\,mag asymmetric oscillation with $\sim$25\,day period. \cite{honeycutt2014} studied these stunted outbursts in more detail.
\paragraph{BZ Cam} The measurements of BZ~Cam showed a brightening event lasting for about 30\,days with 0.7\,mag amplitude and less symmetry.
\paragraph{VZ Scl} The object seem to exhibit a $\sim$1\,mag oscillation.
\paragraph{LX Ser} The system showed an outburst-like event in 2021 lasting for about 7\,months.
\paragraph{CM Del} The light curves of CM~Del reveal a continuous outbursting behaviour with varying amplitude. We also note that some works classified this object as a dwarf nova \citep{ladous1991, lyons2001}.
\paragraph{LQ Peg} A 0.4\,mag outburst appeared on a $\sim$25-day timescale in the photometric observations of LQ Peg.
\paragraph{HS 0506+7725} The light curves of HS~0506+7725 shows a brightening event with $\approx$1\,mag amplitude and lasting for about 40\,days.
\paragraph{LN UMa} Several brightening events occur in the light curves of LN UMa with various shapes and amplitudes. Their duration is typically $20-30$\,days.
 \paragraph{V1082 Sgr} Numerous small amplitude brightenings are revealed by the optical light curves of V1082~Sgr.
 \paragraph{TW Pic} The brightness of the system is increased by $\approx$1\,mag for the period of a few days.
 \paragraph{FY Per} This system shows several stunted outbursts, similarly to V794~Aql. The stunted outbursts of FY~Per are discussed in \cite{honeycutt2001}.

\newpage

\begin{table}[ht!]
\caption{List of VY~Scl type nova-like variables}             
\label{table:vyscl_list}      
\centering                          
\begin{tabular}{l c }        
\hline\hline                 
Object & Reference \\    
\hline                        
HV And    & \cite{henden1995} \\
   & \cite{schwope1992} \\
PX And    & \cite{warner1995} \\
MV Lyr    & \cite{warner1995} \\
V425 Cas  & \cite{warner1995} \\
V751 Cyg  & \cite{warner1995} \\
KR Aur    & \cite{warner1995} \\
V794 Aql  & \cite{warner1995} \\
DW UMa    & \cite{warner1995} \\
LY Hya    & \cite{warner1995} \\
TT Ari    & \cite{warner1995} \\
BZ Cam    & \cite{warner1995} \\
WX Ari    & \cite{warner1995} \\
V442 Oph  & \cite{warner1995} \\
VZ Scl    & \cite{warner1995} \\
BH Lyn    & \cite{warner1995} \\
LX Ser    & \cite{warner1995} \\
CM Del    & \cite{warner1995} \\
VY Scl    & \cite{warner1995} \\
HS 0506+7725 & \cite{downes2001} \\
BH Lyn    & \cite{downes2001} \\
LN UMa    & \cite{downes2001} \\
V504 Cen  & \cite{downes2001} \\
V1082 Sgr  & \cite{downes2001} \\
RX J2337+4308 & \cite{downes2001} \\
TW Pic    & \cite{puebla2007} \\
V533 Her    & \cite{honeycutt2004} \\
FY Per    & \cite{honeycutt2004} \\
LQ Peg    & \cite{honeycutt2004} \\
\hline                                   
\end{tabular}
\end{table}

\begin{figure*}
\centering
\includegraphics[width=\textwidth]{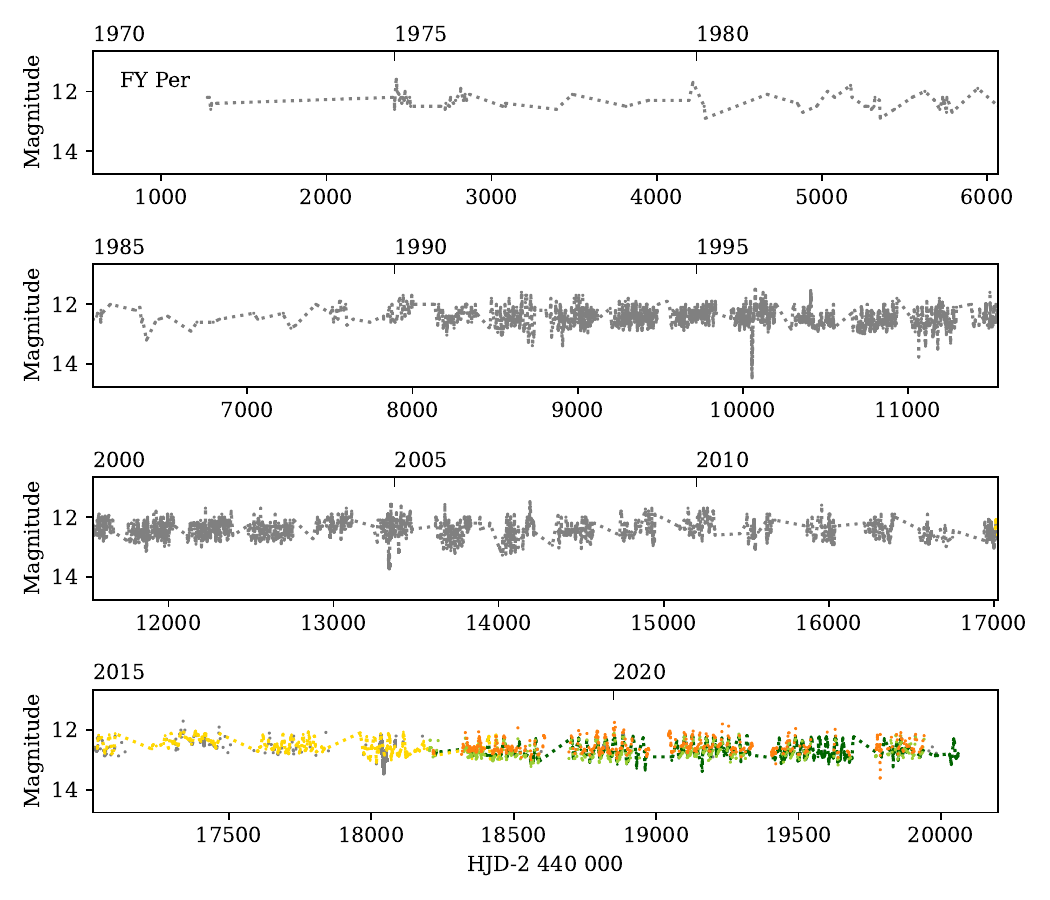}
\includegraphics[width=\textwidth]{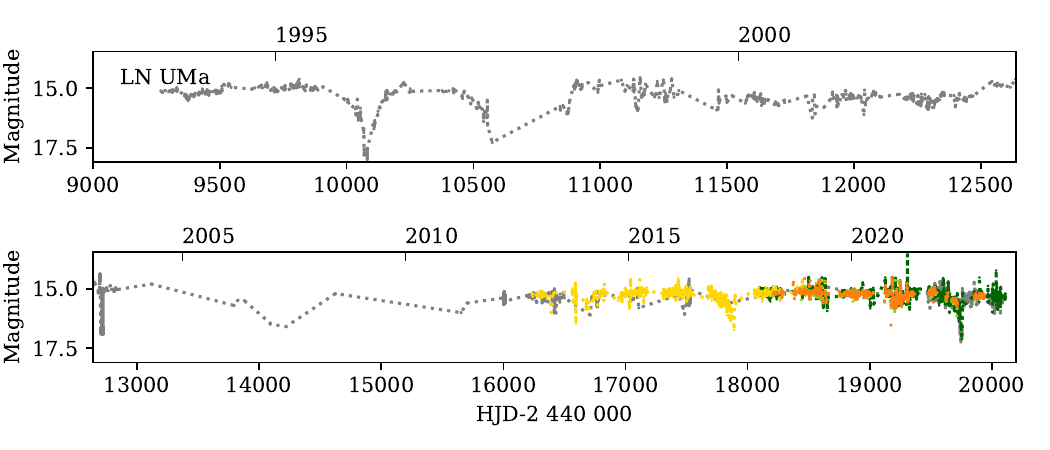}
\caption{Long-term lightcurves of FY~Per and LN~UMa. The top four panels show the available AAVSO (grey), ASAS-SN $V$ (yellow), ASAS-SN $g$ (dark green), ZTF $r$ (orange), ZTF $g$ (light green) light curves for FY~Per. The bottom two panel show the long-term light curves for LN~UMa based on the same catalogues.} 
\label{fig:longterm_lc}
\end{figure*}

\begin{figure*}
\centering
\includegraphics[width=\textwidth]{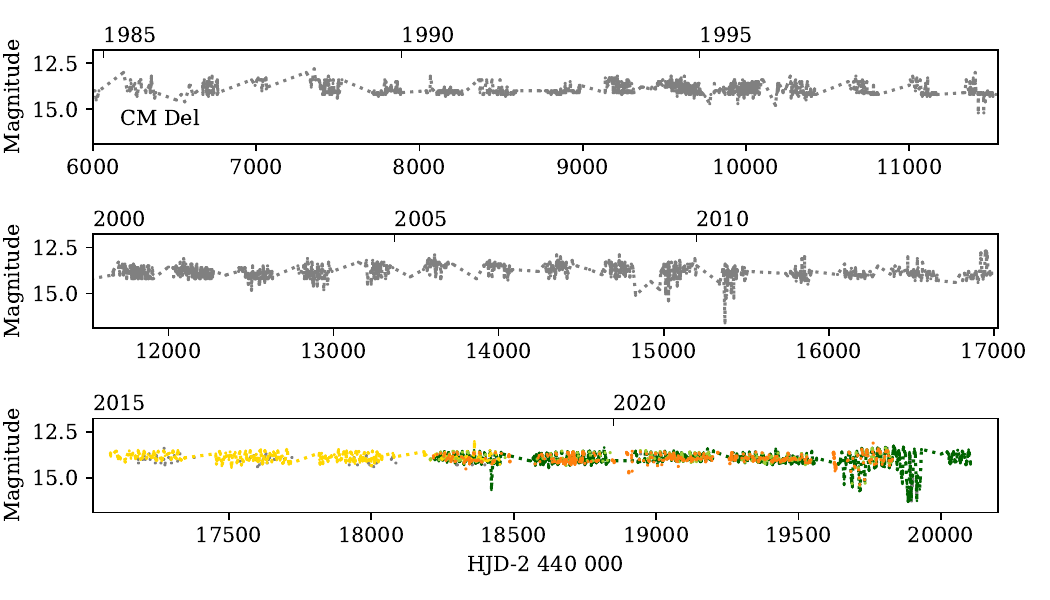}
\includegraphics[width=\textwidth]{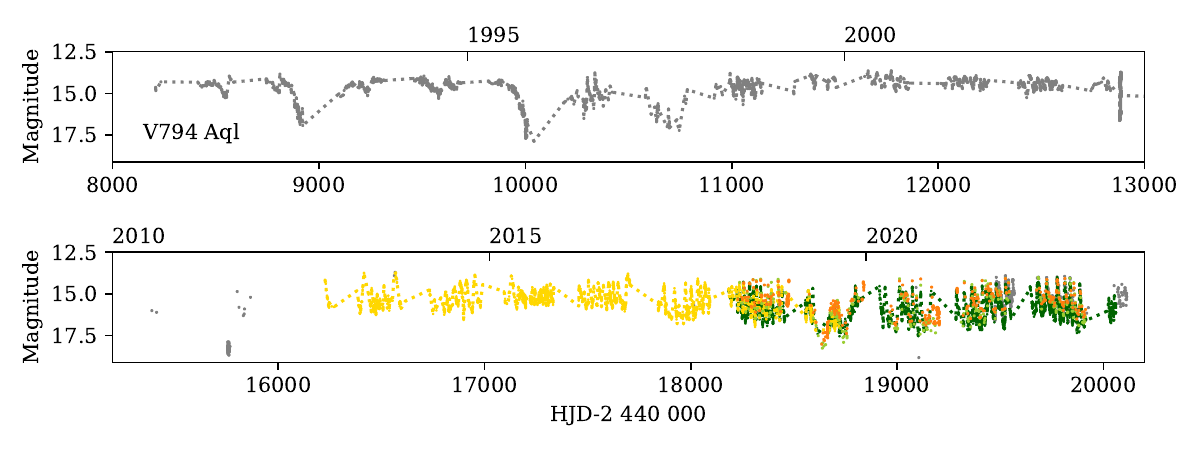}
\caption{Long-term lightcurves of CM~Del and V794~Aql. The top three panels show the available AAVSO (grey), ASAS-SN $V$ (yellow), ASAS-SN $g$ (dark green), ZTF $r$ (orange), ZTF $g$ (light green) light curves for CM~Del. The bottom two panels show the long-term light curves for V794~Aql based on the same catalogues. We note the lack of available data for the latter source between $\sim$2005 and $\sim$2010.}
\label{fig:longterm_lc2}
\end{figure*}

We have inspected the long-term light curves of a few above-mentioned VY Scl type sources using observations from AAVSO, ASAS-SN and ZTF. We show the light curves in Figs.~\ref{fig:longterm_lc} and \ref{fig:longterm_lc2}. For their description, see Sect.~\ref{sect:comparison}.

\begin{figure*}
\centering
\includegraphics[width=0.85\textwidth]{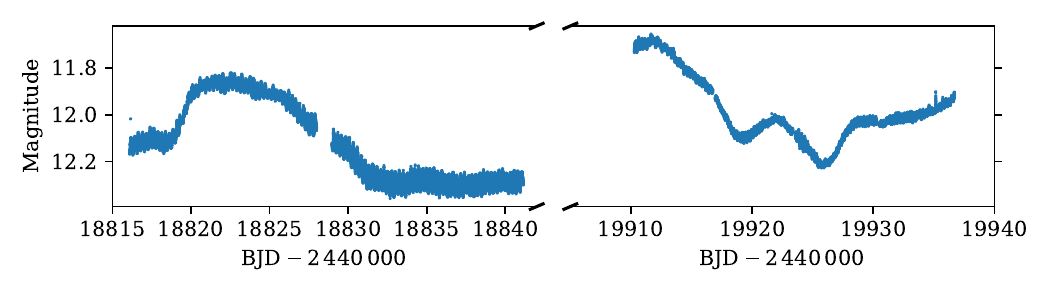}
\caption{High cadence TESS light curve of FY~Per from 2019 and 2022. Both TESS Sectors provided an $\sim$26-day-long light curve with 2-min cadence, which also reveal the $P_{orb} = 6.20\,\textrm{hr}$ \citep{ritter2003} of FY~Per.}
\label{fig:fyper_tess}
\end{figure*}

The Transiting Exoplanet Survey Satellite \citep[TESS,][]{ricker2015} observed FY~Per in 2019 December and in 2022 December with 2-min cadence, providing a $\sim$26-day-long observing sequence for both Sectors. These unprecedentedly high cadence light curves cover one stunted outburst entirely and the decline of presumably another outburst (Fig.~\ref{fig:fyper_tess})
The 2019 measurements reveal the detailed shape of a stunted outburst with fast rise and slower decay.

\FloatBarrier

\onecolumn

\section{Swift UVOT colour-magnitude diagrams}
\label{app:uv_colour}

   \begin{figure*}[ht!]
   \centering
   \includegraphics[width=0.99\textwidth]{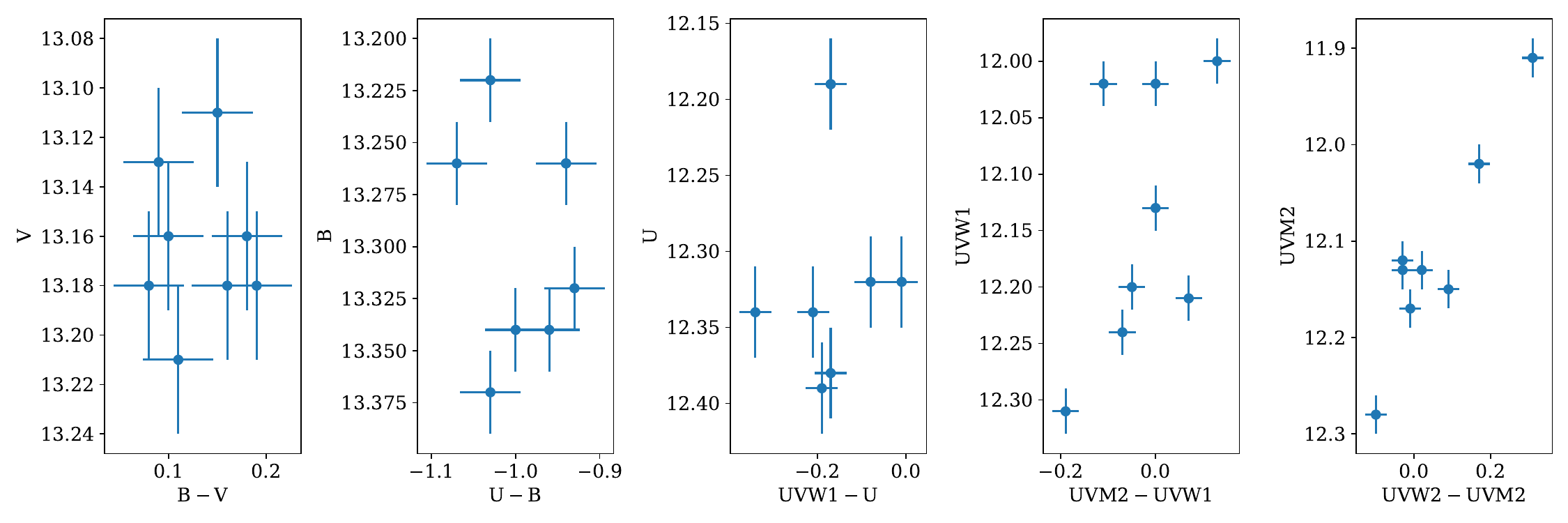}
   \caption{Optical and UV colour-magnitude diagrams of the Swift/UVOT measurements.}
    \label{fig:v704and_swift_cmd}%
    \end{figure*}


The multifilter Swift/UVOT measurements allowed us to obtain colour-magnitude diagrams as well, which are displayed in Fig.~\ref{fig:v704and_swift_cmd}. These represent the high state colour variations of the system.

\end{appendix}

\end{document}